\documentclass[11pt]{article}

\usepackage{amsmath,amssymb}
\usepackage{slashed}
\usepackage{xcolor} %Define colors
\usepackage{graphicx}
\usepackage{url}
\usepackage{cite}

\newcommand{\beq}{\begin{equation}}
\newcommand{\eeq}{\end{equation}}
\newcommand{\bea}{\begin{eqnarray}}
\newcommand{\eea}{\end{eqnarray}}
\newcommand{\nn}{\nonumber \\ }

\newcommand{\hc}{\mathrm{h.c.}}

\newcommand{\cL}{{\cal L}}

\newcommand{\fref}[1]{Fig.~\ref{fig:#1}}
\newcommand{\eref}[1]{Eq.~\eqref{eq:#1}}

\begin{document}

\vspace*{-2cm}
\begin{flushright}
 LPT Orsay 14-26 \\
\vspace*{2mm}
%\today
\end{flushright}

\begin{center}
\vspace*{15mm}

\vspace{1cm}
{\Large \bf
Displaced Vertices from X-ray Lines
} \\
\vspace{1cm}

 {\bf Adam Falkowski${}^{1,*}$, Yonit Hochberg${}^{2,3,\dagger}$, and Joshua T. Ruderman${}^{4,\S}$}

 \vspace*{.5cm}
${}^1$ {\it Laboratoire de Physique Th\'eorique, CNRS -- UMR 8627, \\
 Universit\'e de Paris-Sud 11, F-91405 Orsay Cedex, France} \\
${}^2$ {\it Ernest Orlando Lawrence Berkeley National Laboratory, University of California,\\ Berkeley, CA 94720, USA} \\
${}^3$ {\it Department of Physics, University of California, Berkeley, CA 94720, USA}\\
${}^4$ {\it Center for Cosmology and Particle Physics, \\Department of Physics, New York University,  New York, NY 10003}\\
\vspace*{.5cm}

{\footnotesize
${}^*${\tt adam.falkowski@th.u-psud.fr}\\
${}^\dagger${\tt yonit.hochberg@berkeley.edu}\\
${}^\S${\tt ruderman@nyu.edu}}
\vspace*{0.2cm}

\end{center}

\vspace*{10mm}
\begin{abstract}\noindent
We present a simple model of weak-scale thermal dark matter that gives rise to X-ray lines.
Dark matter consists of two nearly degenerate states near the weak scale, which are populated thermally in the early universe via co-annihilation with slightly heavier states that are charged under the Standard Model. The X-ray line arises from the decay of the heavier dark matter component into the lighter one via a radiative dipole transition, at a rate that is slow compared to the age of the universe.
The model predicts observable signatures at the LHC in the form of exotic events with missing energy and displaced leptons and jets.
As an application, we show how this model can explain the recently observed $3.55$~keV X-ray line.

\end{abstract}

\vspace*{3mm}

\newpage
%%%%%%%%%%%%%%%%%%%%%%%%%%%%%%%%%%
\section{Introduction} \label{sec:intro}

Dark matter in the universe can be indirectly probed by searching for signals of its annihilation or decay.
One promising signature, that may easily stand out from astrophysical backgrounds, is monochromatic photon emission.
Such a signature could arise, for example, due to a two-body decay of dark matter with one or two photons in the final state.
Searches for monochromatic photon lines cover a wide energy spectrum, including the X-ray band from several hundred eV to several hundred keV\@.
Currently operating X-ray telescopes, such as Chandra and XMM-Newton, are providing important limits on the parameter space of dark matter models.
The future ASTRO-H satellite~\cite{Takahashi:2010fa}, scheduled for launch in 2015, should significantly improve sensitivity to both hard and soft X-ray emissions, as well as to soft gamma-rays.

Recently, Refs.~\cite{Bulbul:2014sua, Boyarsky:2014jta} reported detection of  a monochromatic line at $3.55$~keV in galaxy clusters and in the Andromeda galaxy, which may be a signal of dark matter. The line has not been observed by several other studies~\cite{Riemer-Sorensen:2014yda, Jeltema:2014qfa, Malyshev:2014xqa, Anderson:2014tza}, and its significance has been debated in the literature based on different estimates of systematics uncertainties on the X-ray background~\cite{Jeltema:2014qfa, Boyarsky:2014ska,Boyarsky:2014paa}. This detection, although still tentative, highlights the importance of and the opportunities in searching for dark matter via X-rays.

The results from X-ray satellites are most often interpreted as constraints on the parameter space of a sterile neutrino with mass in the  $\sim 1$-$100$~keV range and small mixing with the Standard Model (SM) neutrinos~\cite{Boyarsky:2005us,Boyarsky:2006zi, Boyarsky:2006fg, Abazajian:2006yn, Watson:2006qb, Boyarsky:2006ag, Abazajian:2006jc, Watson:2011dw}.
The observed 3.55~keV line can be accommodated by decays of sterile neutrino dark matter~\cite{Bulbul:2014sua, Boyarsky:2014jta}, assuming a very large initial lepton asymmetry relative to the baryon asymmetry, $\Delta L \sim 10^6 \, \Delta B$~\cite{Abazajian:2014gza}.
However, sterile neutrino dark matter is by no means the unique explanation.
Indeed, the detection of the X-ray line prompted construction of a wider variety of models capable of explaining such a feature; see {\it e.g.}~\cite{Finkbeiner:2014sja, Higaki:2014zua, Jaeckel:2014qea, Lee:2014xua, Krall:2014dba, Aisati:2014nda, Frandsen:2014lfa, Nakayama:2014ova, Choi:2014tva, Cicoli:2014bfa, Kolda:2014ppa, Bomark:2014yja, Liew:2014gia, Nakayama:2014cza, Ko:2014xda, Demidov:2014hka, Dudas:2014ixa, Babu:2014pxa, Lee:2014koa, Baek:2014poa,Abada:2014zra,Chiang:2014xra,Cline:2014kaa,Henning:2014dha,Boddy:2014qxa}.
In this paper we discuss the possibility that an X-ray line arises from weak-scale dark matter with a thermal history.
Our philosophy is to provide a UV complete model with minimal ingredients, in the spirit of minimal dark matter~\cite{Cirelli:2005uq}.
Within such a framework we can discuss possible links between the X-ray line and collider phenomenology.

\begin{figure}[t]
    \begin{center}
         \includegraphics[width=0.8\textwidth]{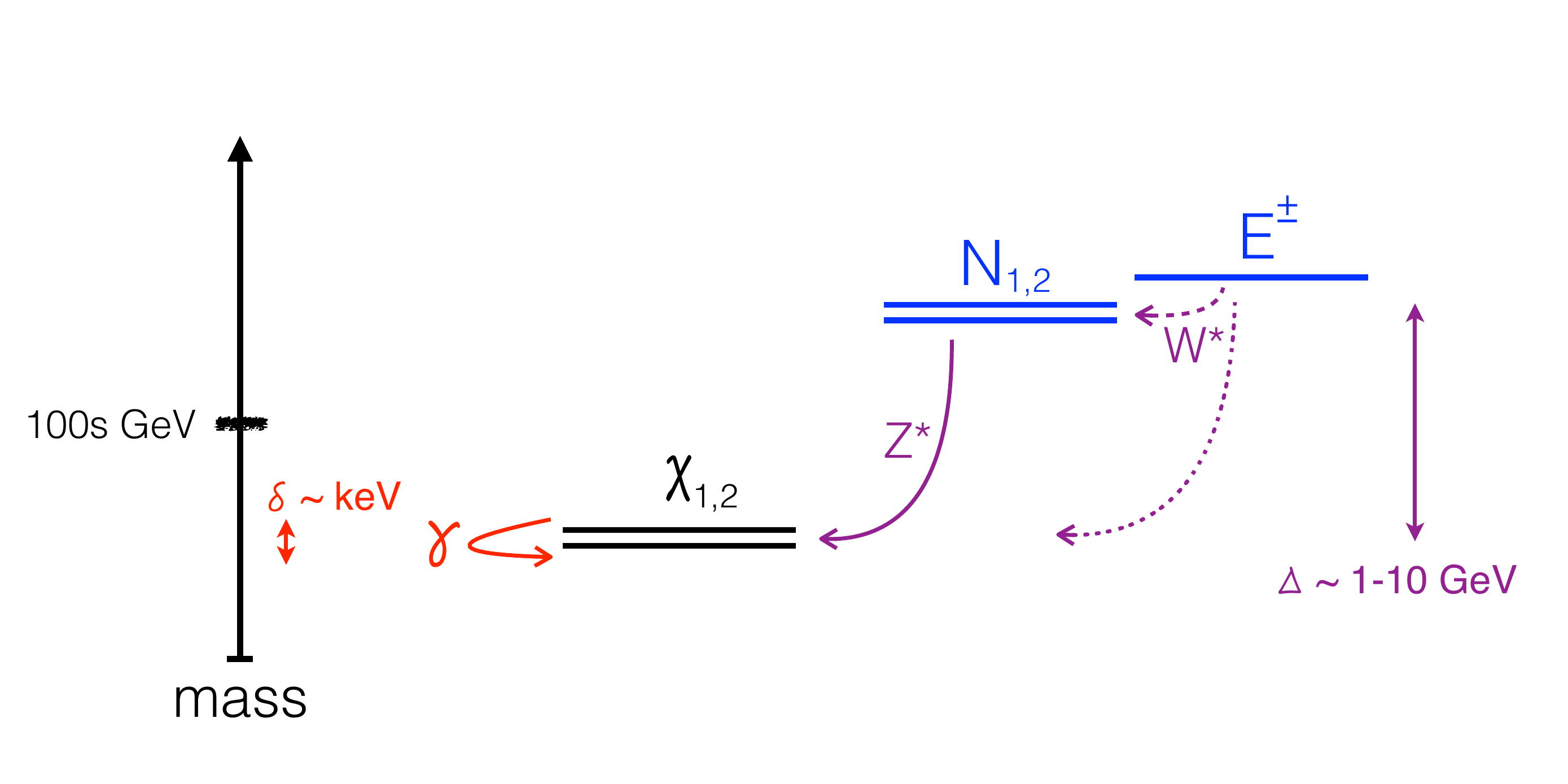}
\vspace*{-2mm}
          \caption{\footnotesize
          A schematic description of the model. Dark matter is comprised of two nearly degenerate states $\chi_{1,2}$ which are both populated thermally in the early universe via co-annihilation with a vector-like doublet. All states have masses of order the weak scale. X-ray lines arise from the radiative dipole transition between the two dark matter states.
          The charged and neutral components of the doublet, $E^{\pm}$ and $N_{1,2}$, can be produced at the LHC, followed by decay to dark matter and an off-shell $Z$ or $W$ boson, leading to distinct signatures.
          }
\label{fig:schema}
\end{center}
\vspace*{-3mm}
\end{figure}

The basic properties of our setup are the following:
\begin{enumerate}
\item Dark matter consists of two weak-scale states separated by a small mass splitting;
\item Both dark matter states are populated thermally in the early universe;
\item  The X-ray line arises through a radiative decay of the heavier dark matter component into the lighter one.
\end{enumerate}
More precisely, dark matter in our model is a vector-like singlet fermion.
We also introduce a vector-like fermion doublet charged under the SM electroweak group, which plays the role of the link between  dark matter and the SM sector.
The singlet interacts with the doublet via Yukawa couplings, leading to a mixing between the dark matter and the active states after electroweak symmetry breaking.
This mixing has several implications.
First, dark matter can be thermally populated in the early universe via co-annihilation with the active states.
Second, the mixing induces a splitting between the two Majorana eigenstates of dark matter.
The splitting is controlled by the magnitude of the Yukawa couplings, which allows one to obtain ${\cal O}$(keV) splitting in a technically natural way.
Finally, dipole transitions between the two dark matter states are generated by one-loop diagrams with the charged component of the doublet in the loop.
Therefore, all the ingredients required to produce an X-ray line are present in this model.
As an application, we describe how this setup can accommodate the recently observed $3.55$~keV line.
The model furthermore links X-ray lines to collider phenomenology.
The doublet can be produced with a significant cross section at the LHC, and subsequently decays to dark matter and an off-shell $Z$ or $W$ boson.
It turns out that, in the parameter space relevant for X-ray lines, the doublet lifetime is in the millimeters to meters ballpark.
Therefore the  model predicts distinct signatures at the LHC: leptons and jets with displaced vertices accompanied by missing energy. A schematic description of the model is presented in Fig.~\ref{fig:schema}.

Scenarios with X-ray lines generated by transitions among weak scale dark matter states have also been proposed by Refs.~\cite{Finkbeiner:2014sja,Ko:2014xda,Lee:2014koa,Chiang:2014xra}.  Ref.~\cite{Finkbeiner:2014sja} considers an effective theory of a weak scale state with cosmologically long lifetime, and our model can be viewed as a UV completion and cosmological history for this scenario.  The related models of Refs~\cite{Ko:2014xda,Lee:2014koa,Chiang:2014xra} have different field content than our model, and in particular all include extra scalar fields at the weak scale, beyond the Higgs boson.  Our model extends the Standard Model by only fermionic fields, factoring the dark matter sector from the solution to the hierarchy problem, and has distinct collider phenomenology.

The organization of this paper is as follows. In Section~\ref{sec:model} we present the model and discuss the spectrum and symmetries (Section~\ref{ssec:sym}), dark matter decays (Section~\ref{ssec:decay}) as well as the relic abundance (Section~\ref{ssec:abundance}). Section~\ref{sec:bulbuline} analyzes the parameter space relevant for the observed $3.55$~keV line and the resulting LHC phenomenology. In Section~\ref{sec:anyother} we describe the general parameter space of X-ray and gamma-ray lines within our model. We conclude in Section~\ref{sec:conclusions}.

%%%%%%%%%%%%%%%%%%%%%%%%%%%%%%%%%%%%%%%%%%%
\section{Model} \label{sec:model}

We extend the SM by introducing the following new fermionic fields:
\begin{itemize}
\item  A vector-like fermion pair $\chi$, $\chi^c$ that is a singlet under the SM gauge group;
\item A  vector-like pair of doublets $L = (N,E)$ and $L^c=(N^{c},E^c)$  transforming as $(1,2)_{-1/2}$ and $(1,\bar 2)_{+1/2}$ under the SM gauge group.
\end{itemize}
In our notation, each of the above fields is a 2-component spinor, and we follow the conventions of Ref.~\cite{Dreiner:2008tw}.
No other new particles are necessary for our purpose.
We assume that these fields carry no lepton number.
As a result, they do not couple to the SM leptons nor mix with them.
Therefore, the only way the new sector interacts with the SM is via the $SU(2)_L \times U(1)_Y$ gauge interactions.

%---------------------------------------------------
 \subsection{Spectrum and symmetries}\label{ssec:sym}

The potential of the new physics sector of our model is given by
\beq
\label{eq:spectrum}
- \cL = M_\chi \chi \chi^c  + M_L L L^c +   y \chi  H L +  \tilde y \chi^c  H^\dagger L^c +  \lambda \chi H^\dagger L^c +  \tilde \lambda  \chi^c  H L +   \hc,
\eeq
where $H$ is the SM Higgs field with the vev $\langle H \rangle = (0,v/\sqrt 2)$ and $v=246$~GeV\@.
From our point of view, the interesting part of the parameter space will be that where  $M_L \gtrsim M_\chi$ and the vector-like masses are of order the weak scale. We have included the Yukawa couplings between the singlet fermions, doublet fermions, and Higgs boson. Once the Higgs obtains a vev, the singlet fermions mix with  the neutral components of the fermion doublets. For simplicity we take the Yukawa couplings to be real, and choose the phases of $L^c$ and $\chi^c$ such that $M_L$ and $M_\chi$ are real and positive. We will assume the Yukawa couplings are small, and treat the mixing as a perturbation.
This assumption is technically natural. Indeed, the new physics sector has a $U(1)^4$ chiral symmetry, which is softly broken to $U(1)_\chi \times U(1)_L$ by the vector-like mass terms. Each of the Yukawa coupling  breaks one of the $U(1)$ symmetries.
Thus, the symmetry is enhanced in the limit where any Yukawa coupling vanishes.

We can envisage two discrete symmetries acting on the new physics sector, in analogy to `messenger parity' of gauge mediation\cite{Dimopoulos:1996ig}:
\begin{itemize}
    \item {\em L-parity}, which interchanges $L \leftrightarrow L^c$, and
    \item {\em $\chi$-parity}, which interchanges $\chi \leftrightarrow \chi^c$.
\end{itemize}
The former implies $y = \lambda$ and $\tilde y =  \tilde \lambda$, while the latter implies $y = \tilde \lambda$ and $\tilde y =  \lambda$.
These symmetries will play an important role later on, because we will find that observable X-rays occur within the parameter space with approximately degenerate Yukawa couplings.
$L$-parity is broken by the coupling to the hypercharge gauge boson, since in the SM left-handed and right-handed fermions have different $U(1)_Y$ charges.
Thus, starting from $y= \lambda$ at some scale, the renormalization group running will induce a splitting at the 2-loop level, {\it e.g.} from a top loop connected to a Higgs and $Z$-boson which are exchanged between the external singlet and doublet fermions. For this reason, degeneracy smaller than $|1 - \lambda /y|  \lesssim y_t^2 g^2/(4\pi)^4\sim 10^{-4}-10^{-5}$ (where $y_t$ is the top Yukawa coupling and $g$ the electroweak gauge coupling) requires fine-tuning. On the other hand, $\chi$-parity is exact, and therefore $|1 - \tilde \lambda /y|$ can be arbitrarily small without fine-tuning.

In general, the eigenstates of the mass matrix are Majorana-type, as soon as both $U(1)_\chi$ and $U(1)_L$ numbers are broken by the Yukawa interactions,  that is
if  ($\lambda$ or $ \tilde \lambda$) and ($y$ or  $\tilde y$) are non-zero.
We denote these mass eigenstates as $\chi_1$, $\chi_2$, $N_1$, $N_2$.
In the limit $v \to 0$ the de-facto Dirac eigenstates can be recast in the Majorana form as $\chi_{1,2} = {\chi  \pm \chi^c \over \sqrt 2}$ with mass $M_\chi$, and $N_{1,2} = {N \mp N^c \over \sqrt 2}$  with mass $M_L$.\footnote{%
In our conventions, the mass parameters of $\chi_2$ and $N_1$ in the Lagrangian are negative.
This is taken into account in the Feynman rules.
Alternatively, one can rescale $\chi_2 \to i \chi_2$ and $N_1 \to i N_1$ and work with positive mass terms. In what follows, masses $m_{N_{1,2}}$ and $m_{\chi_{1,2}}$ will always implicitly refer to absolute values of masses.
 }
For  $v > 0$,  the eigenstates $\chi_{1,2}$ acquire small active components $N$ and $N^c$:
\bea
\label{eq:rotation}
\chi_1 & \approx  &{1  \over \sqrt 2}  \left ( \chi+ \chi^c +  s_{11} (N + N^c) +  s_{12} (N - N^c) \right ),
\nn
\chi_2 &\approx&  {1  \over \sqrt 2}  \left ( \chi - \chi^c +  s_{21} (N + N^c) +  s_{22} (N - N^c) \right ),
\eea
and the eigenstates $N_{1,2}$ become
\bea\label{eq:N12rot}
N_1 &\approx&  {1  \over \sqrt 2}  \left ( N - N^c -  s_{12} (\chi + \chi^c) -  s_{22} (\chi - \chi^c) \right ),
\nn
N_2 & \approx  &{1  \over \sqrt 2}  \left ( N+ N^c -  s_{11} (\chi + \chi^c) -  s_{21} (\chi - \chi^c) \right ).
\eea
The mixing angles $s_{ij}$ are given by
\bea
\label{eq:sij}
s_{11} &=&  -  {v \over 2 \sqrt 2 (M_L - M_\chi)} \left (y + \tilde y + \lambda + \tilde \lambda \right ),
\nn
s_{12} &=&   {v \over 2 \sqrt 2 (M_L + M_\chi)} \left (y -  \tilde y -  \lambda + \tilde \lambda \right ),
\nn
s_{21} &=&  -  {v \over 2 \sqrt 2 (M_L + M_\chi)} \left (y -  \tilde y +  \lambda - \tilde \lambda \right ),
\nn
s_{22} &=&   {v \over 2 \sqrt 2 (M_L - M_\chi)} \left (y +  \tilde y -  \lambda -  \tilde \lambda \right ).
\eea
The mixing is suppressed by the magnitude of the Yukawa couplings.
Further suppression may arise thanks to discrete symmetries in the new physics sector.
In particular, in the $L$-parity limit only the symmetric combination of $N+N^c$ couples to dark matter,
thus $N_1$ does not mix into $\chi_{1,2}$.
Similarly, in the $\chi$-parity limit only $\chi + \chi^c$ has a Yukawa coupling to the active sector,
 therefore $\chi_2$ does not mix with $N$, $N^c$.
On the other hand, for $M_L \approx M_\chi$, some mixing angles may be enhanced by $M_\chi/\Delta$, where $\Delta \equiv M_L - M_\chi$.
We shall see that obtaining a thermal relic abundance of dark matter dictates $\Delta/M_\chi \sim 10^{-1}-10^{-2}$, and that fitting the $3.55$~keV line will require degeneracy amongst Yukawa couplings at the percent to per mille level.
(For a general X-ray line, the amount of degeneracy amongst the couplings can vary depending on the observed flux; see Section~\ref{sec:anyother}.)
This will lead to the following hierarchy of the mixing angles:
\bea
{\rm \bf L-parity:} & \quad & |s_{11}| \gg |s_{21}| > |s_{22}| \gg  |s_{12}|,
\nn
{\rm \bf \chi-parity:} & \quad & |s_{11}| \gg |s_{12}| > |s_{22}| \gg  |s_{21}|.
 \eea

When both $U(1)_\chi$ and $U(1)_L$ are broken, the mixing induces a small splitting $\delta$ between the quasi-degenerate states.
It  arises at the quadratic order in Yukawa couplings, and we find
\beq
\label{eq:delta}
\delta \equiv m_{\chi_2}  - m_{\chi_1}  \approx   {v^2 \over  M_L^2- M_\chi^2} \left [
M_L \left ( y \lambda + \tilde y \tilde \lambda \right )   +  M_\chi \left ( y \tilde  \lambda + \tilde y \lambda \right ) \right ].
\eeq
In what follows, the states $\chi_2$ and $\chi_1$ will be the particles that account for dark matter in the universe, and the transition between these two states will give rise to the X-ray line. Then, for $M_L$ and $M_\chi$ at the weak scale, we will see that the Yukawa couplings must be of the order of $10^{-5}$ up to $10^{-3}$ for $\delta$ between a keV and an MeV\@.
This implies the mixing angles are small,  and the perturbative expansion of the eigenstates provides a very good approximation of the true spectrum.
The heavier neutral states are split by a similar amount $m_{N_2} - m_{N_1} \approx \delta$, but since this plays no phenomenological role we will always approximate $m_{N_i} \approx M_L$. We consider only $M_L\gtrsim M_\chi$ since for mass splittings relevant for X-ray lines, dark matter that is doublet-like is already excluded by direct detection bounds~\cite{Cirelli:2005uq, Essig:2007az}.

 In general, Yukawa couplings of ${\cal O}(1)$ would give rise to weak scale splitting for weak-scale dark matter. The splitting $\delta$ of Eq.~\eqref{eq:delta} gives rise to X-rays only if at least one Yukawa is small. This smallness can originate from breaking at a high scale of the $U(1)^4$ symmetry in the dark sector, in a similar manner to the methods that address the flavor puzzle of the SM (see {\it e.g.}~\cite{Froggatt:1978nt, Leurer:1992wg, Leurer:1993gy, ArkaniHamed:1999dc, Gherghetta:2000qt, Nelson:2000sn, Huber:2000ie}).

%---------------------------------------------------------
\subsection{Dark matter decays}\label{ssec:decay}

Thanks to the mixing with the active states, $\chi_1$ and $\chi_2$ acquire a small coupling to the SM gauge fields.
Furthermore, the mass splitting between $\chi_2$ and $\chi_1$ opens the phase space for the decays  $\chi_2 \to \chi_1 \nu \nu$ and $\chi_2 \to \chi_1 \gamma$.
The former is a 3-body decay process proceeding at tree-level via an off-shell $Z$ boson.
We find
\beq
\Gamma (\chi_2  \to \chi_1 \nu \nu) \approx
 {3 \delta^5 \over 40 \pi^3 v^4}   \left (s_{12} s_{21} + s_{11} s_{22} \right )^2\,.
\eeq
The latter decay process arises at one-loop level, via dipole diagrams with the charged component of the vector-like lepton doublet and the SM $W$-boson in the loop.
We find (see~\cite{Haber:1988px} for a useful gauge choice)
\beq\label{eq:n2width}
\Gamma (\chi_2  \to \chi_1 \gamma) =  { e^2 \over 16  \pi^5}  {\delta^3 \over M_L^2}   \left  [  c_1 I_1   +    c_2 I_2   \right ]^2\,,
\eeq
where $e$ is the electromagnetic gauge coupling, and the integrals are defined as
\bea
I_{1} &=&   M_L^2  \int_0^1 ds  { s (2 s - 1) \over s m_W^2 + (1- s) M_L^2 - s (1-s) M_\chi^2}\,,
\nn
 I_{2} &=&  M_L^2  \int_0^1 ds  {1-s \over s m_W^2 + (1- s) M_L^2 - s (1-s) M_\chi^2}\,,
\eea
and they are ${\cal O}(1)$ in the interesting parameter region $M_L \sim M_\chi \sim m_W$.
The coupling combinations that enter the 2-body width in Eq.~\eqref{eq:n2width} are given by
\bea
c_1 &= &
{m_W^2 \over (M_L^2 - M_\chi^2)^2 } \left [ (M_L^2 + M_\chi^2) (y \tilde y - \lambda \tilde \lambda) + M_L M_\chi (y^2 + \tilde y^2 - \lambda^2 - \tilde \lambda^2) \right ]\,,
\nn
c_2 & = &   {y \tilde y - \lambda \tilde \lambda   \over 4}\,.
\eea

We note that the 3-body width is  proportional to   $\delta^5$.
As a result, in the interesting parameter space where $\delta$ is small,  the 3-body decay is completely subdominant to the 2-body radiative decay width, which is proportional to  $\delta^3$.
In addition, notice that the $\chi_2 \to \chi_1$ decay width (both 2- and 3-body) vanishes both in the $L$-parity and in the $\chi$-parity limit.
In the $\chi$-parity limit, $\chi_2$ is charged under the $\chi\leftrightarrow \chi^c$ symmetry while $\chi_1$ is not, and so any $\chi_2\to \chi_1$ transition is forbidden. In the $L$-parity limit, we can understand the vanishing width as follows.  Because $\chi$ and $\chi^c$ are singlets, the dipole operator is proportional to the dipole of the active states, $N_1 \sigma_{\mu \nu} N_2 F^{\mu \nu} + \hc$. The analogous operators with only $N_1$ or $N_2$ vanish due to spinor anti-commutation. Since in the $L$-parity limit $N_1$ does not mix with or couple to the dark matter, the dipole operator vanishes. This holds to all loop orders. The 3-body decay vanishes in the $L$-parity limit due to the vanishing coupling to the $Z$ boson.

%---------------------------------------------------------
\subsection{Relic Abundance}\label{ssec:abundance}

We now consider the relic abundance of dark matter in our model.  It is not possible for $\chi$ to have the correct abundance from freezeout of $\chi \chi$ annihilations.  Given the small size of  $\chi-L$ mixing required for an observable X-ray line, the $\chi \chi$ annihilation cross section is too small and $\chi$ would overclose the universe.
Fortunately, the required dark matter abundance can result from co-annihilation~\cite{Griest:1990kh} between $\chi$ and $L$, as long as their mass difference is of the order $10$~GeV.

Let us quickly review the parametric dependence of the dark matter abundance in the co-annihilation case~\cite{ArkaniHamed:2006mb}.
The relic abundance is given by
\beq
\Omega_{\rm DM} h^2 = {8.7 \times 10^{-11} {\rm GeV}^{-2} \over \sqrt{g_*}  \int_{x_f}^\infty dx \langle \sigma_{\rm eff} v \rangle x^{-2}}\,,
\eeq
where $g_*$ is the effective number of relativistic degrees of freedom at freeze-out,
$x \equiv m_{\rm DM}/T$ and $x_f \approx 22$ depends on the freeze-out temperature.
The effective cross section is a weighted average of the annihilation cross sections of the co-annihilating particles,
\beq
\label{eq:coanal}
\langle \sigma_{\rm eff} v \rangle  = {\sum_{i j} \sigma_{i j} w_i w_j \over \left ( \sum_i w_i \right )^2 },
\qquad  \langle \sigma_{ij} v \rangle =  \sigma_{i j} x^{-n},
\qquad w_i = \left ( m_i \over m_{\rm DM} \right )^{3/2} \exp \left [ -x (m_i/m_{\rm DM} -1) \right ]
\eeq
where the index $i$ runs over co-annihilating particles, $m_1 \equiv m_{\rm DM}$, and $n=0(1)$ for $s(p)$-wave annihilation. Co-annihilation between $\chi$'s and $L$'s is effective when they are in thermal equilibrium, which holds as long as the mixing angles of Eq.~\eqref{eq:sij} obey ${\rm max}(s_{ij})\gtrsim (x_f \sqrt{g_*}M_L/M_{\rm Pl})^{1/2}$~\cite{ArkaniHamed:2006mb}, with $M_{\rm Pl}$ the Planck mass. In the parameter space of interest, this requires Yukawa couplings $\gtrsim 10^{-5}$.

In our case, only the active states $L$ have an appreciable annihilation cross section, and we find
\bea
\label{eq:LL}
\langle \sigma_{\rm eff} v \rangle &=&  {  \sigma_{LL}  w (x)^2 \over \left (w(x) + 1/2 \right )^2 },
\nn
w(x) &=  &  \left ( 1 + {\Delta  \over M_\chi} \right )^{3/2} \exp \left ( - { x \Delta \over M_\chi}\right ),
\nn
\sigma_{LL}  &=&  {81 g_L^4 + 12  g_L^2 g_Y^2 + 43 g_Y^4 \over 2048   \pi M_L^2 },
\eea
where $\Delta = M_L - M_\chi$, and $g_L$ and $g_Y$ are the weak and hypercharge gauge couplings.
Clearly, $\Delta/M_\chi$  must be comparable to $1/x_f \sim 0.1$ for the co-annihilation to be efficient.
In \fref{coan} we plot $\Delta$ as a function of $M_\chi$, for which the relic  abundance of $\chi_1$ and $\chi_2$ matches the observed dark matter abundance.
The relic density was obtained numerically with {\tt micrOMEGA}~\cite{Belanger:2006is}, using a custom input model exported to {\tt CalcHEP}~\cite{Belyaev:2012qa} from {\tt FeynRules} \cite{Alloul:2013bka}.
We  then verified that the analytic approximation in  \eref{coanal} and \eref{LL} reproduces well the numerical results.
We see that the allowed mass range for $M_L$ and $M_\chi$ is severely constrained.
The shaded region to the left corresponds to $M_L \lesssim 100$~GeV, making it accessible at LEP-2.
The shaded region on the right represents the upper limit $M_L \approx 700$~GeV, as we impose that the lightest state is singlet-like. (As mentioned earlier, for splitting $\delta$ producing X-ray lines, doublet-like dark matter is excluded by direct detection searches~\cite{Cirelli:2005uq, Essig:2007az}.)
For a given $M_\chi$, the mass difference $\Delta$ is uniquely determined, and spans the range between 0~and~11~GeV\@.
In this computation, we assumed that  kinetic equilibrium between $\chi$ and $L$ is maintained until the decoupling via $\chi L \to \chi L$ scattering, which indeed holds in the parameter space of interest.

\begin{figure}[t]
    \begin{center}
         \includegraphics[width=0.7\textwidth]{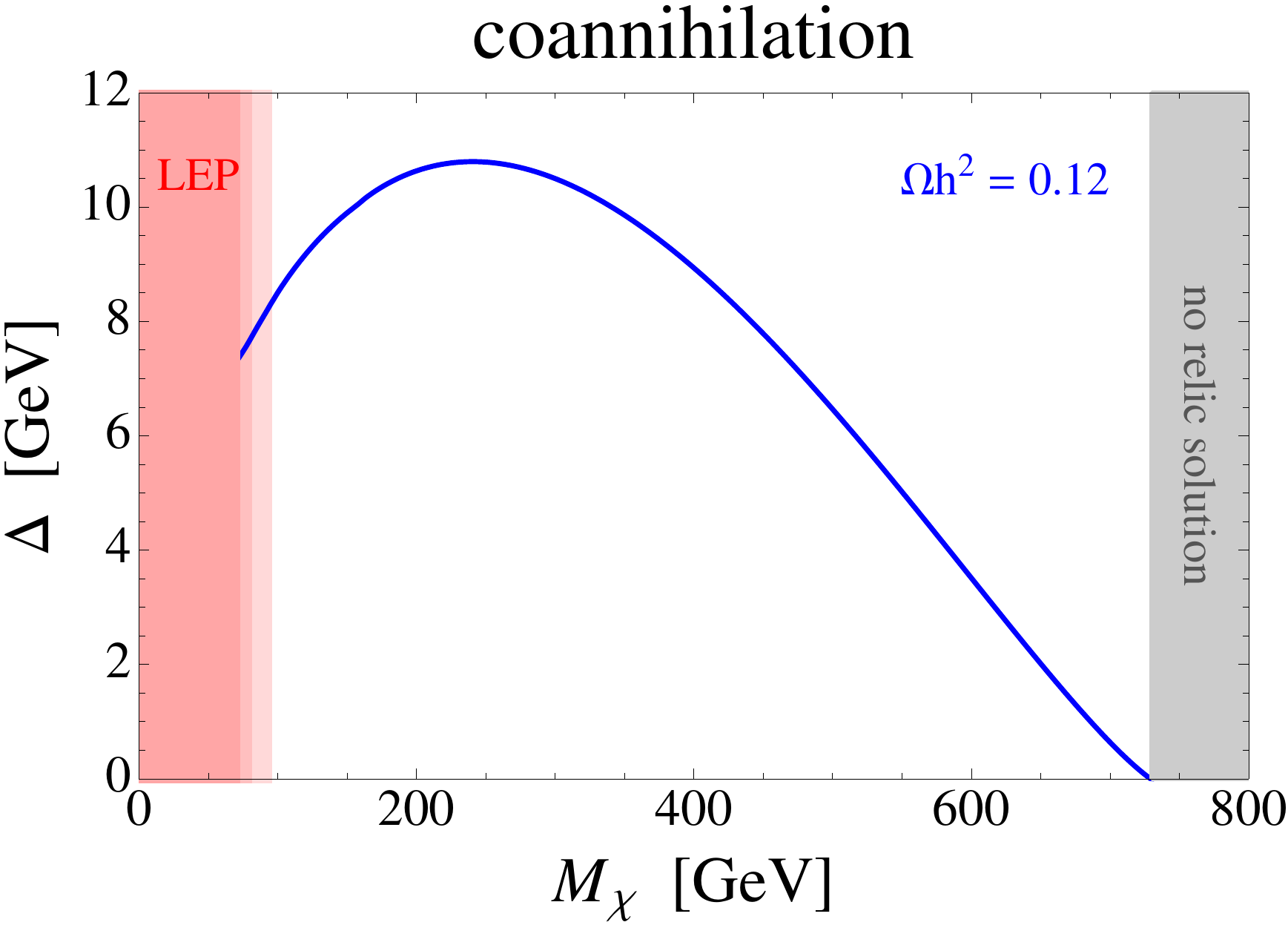}
\vspace*{-2mm}
          \caption{\footnotesize
          The splitting $\Delta = M_L - M_\chi$ between the active doublet $L$ and the dark matter states $\chi$ required to obtain the observed dark matter relic abundance via thermal co-annihilation between $\chi$ and $L$.
 The shaded region to the left is where $M_L \lesssim 100$~GeV, making it accessible at LEP-2. The shaded region on the right is where there is no relic solution for $\chi$.
   }
\label{fig:coan}
\end{center}
\vspace*{-3mm}
\end{figure}

%%%%%%%%%%%%%%%%%%%%%%%%%%%%%%%%%%%%%%%%%%%
\section{3.55 keV line} \label{sec:bulbuline}

The setup described above serves as a thermal weak-scale dark matter module for X-ray emission. Here we demonstrate this by fitting to the recently reported $3.55$~keV line~\cite{Bulbul:2014sua, Boyarsky:2014jta}. The exercise is instructive irrespective of the controversy surrounding this particular observation; a discussion of the general parameter space of X-rays follows in Section~\ref{sec:anyother}.

%-------------------------------------------------
\subsection{Parameter space}

Our model has 6 new parameters: two vector-like masses $M_L$ and $M_\chi$, and four Yukawa couplings $y$, $\tilde y$, $\lambda$, $\tilde \lambda$.
Requiring that the observed dark matter abundance arises from thermal co-annihilation between $\chi$ and $L$ fixes $M_L$ for a given $M_\chi$, and implies
$100~{\rm GeV} \lesssim M_\chi \lesssim 700$~GeV\@.
Now, in order to fit the position and the flux of the X-ray line observed in Refs.~\cite{Bulbul:2014sua, Boyarsky:2014jta}, we impose two additional constraints:
\beq
\label{eq:match}
\delta=m_{\chi_2} - m_{\chi_1} \approx 3.55~{\rm keV},  \qquad
\Gamma (\chi_2  \to \chi_1 \gamma)   \approx 3.1 \times 10^{-47} M_\chi.
\eeq
The decay width is obtained translating the best fit parameters in the 7.1~keV sterile neutrino model quoted in Ref.~\cite{Bulbul:2014sua}.
The dependence on $M_\chi$  arises via the number density of dark matter: as the mass increases the number density decreases, therefore a faster decay is needed for a fixed flux of the X-ray line.
One should also take into account that in our case $\chi_2$ accounts for only half of the dark matter abundance.
The width in \eref{match} corresponds to the lifetime $\tau \approx  {500~{\rm GeV} \over M_\chi} \cdot 10^{12}~{\rm years}$, which is much longer than the age of the universe for $M_\chi$ at the weak scale.

The conditions in \eref{match} fix another two parameters, for example two of the four Yukawa couplings.
From now on, we consider the subset of the parameter space where only two Yukawa couplings: $y$, and $\lambda$ or $\tilde \lambda$, are non-zero.\footnote{%
If only $y$ and $\tilde y$ are non zero then $m_{\chi_2} - m_{\chi_1} = 0$ because the $U(1)_\chi$ symmetry is unbroken.}
In such a  case, for a given dark matter mass $M_\chi$,  all other parameters are determined by fitting the relic abundance and the X-ray line properties.\footnote{%
For a given splitting $\delta$ and decay width $\Gamma$ there are several equivalent solutions that differ by interchanging the two Yukawa couplings or flipping their signs. For definiteness, we always pick the solution where both Yukawas are positive and $y$ is the larger coupling.}
It is convenient to define the level of degeneracy required between the Yukawa couplings,
\beq\label{eq:eps}
\epsilon\equiv \left\{
     \begin{array}{ll}
         y/\lambda-1 & \quad{\rm for}\ \tilde y = \tilde \lambda =0 \\
         y/\tilde \lambda-1 & \quad{\rm for}\ \tilde y = \lambda =0
     \end{array}
   \right.
\eeq
which indicates the size of deviation from the $L$-parity and $\chi$-parity limits in each case respectively.
We do not find any qualitatively new phenomena if we allow all four Yukawa couplings to be simultaneously non-zero.

In \fref{couplings} we show the required size of the Yukawa couplings as a function of $M_\chi$ that satisfy the conditions Eq.~\eqref{eq:match}. (The results are depicted for both $\tilde y=\tilde \lambda=0$ and $\tilde y=\lambda=0$, exhibiting similar behavior. The reason is that between these two cases, the two-body decay rate is symmetric under $\lambda \leftrightarrow \tilde\lambda$, and the mass splitting $\delta$ is symmetric up to a factor of $M_L/M_\chi$ which is close to unity, leading to nearly identical results.)
We find that such Yukawa couplings have to be of order  $10^{-5}$ to reproduce the splitting $\delta$ via Eq.~\eqref{eq:delta}.
Furthermore,  ${\cal O}(1\%)$  degeneracy between $y$ and $\lambda$ (or $y$ and $\tilde \lambda$) is required to fit $\Gamma (\chi_2  \to \chi_1 \gamma)$ via Eq.~\eqref{eq:n2width}.
In other words,  for $y$ and $\lambda$ non-zero, the Yukawa couplings need to be approximately  $L$-parity symmetric, otherwise the radiative decay $\chi_2 \to \chi_1 \gamma$ occurs too fast and is excluded by X-ray limits.
Analogously, in the case  when $y$ and $\tilde \lambda$ are non-zero, these two have to be approximately degenerate, that is to say, approximately $\chi$-parity symmetric.
Similar conclusions are obtained for more general Yukawa couplings: to match the observed energy and flux of the $3.55$~keV X-ray line, Yukawa couplings close to the $L$-parity or close to the $\chi$-parity limit are required.

\begin{figure}[t]
    \begin{center}
         \includegraphics[width=1\textwidth]{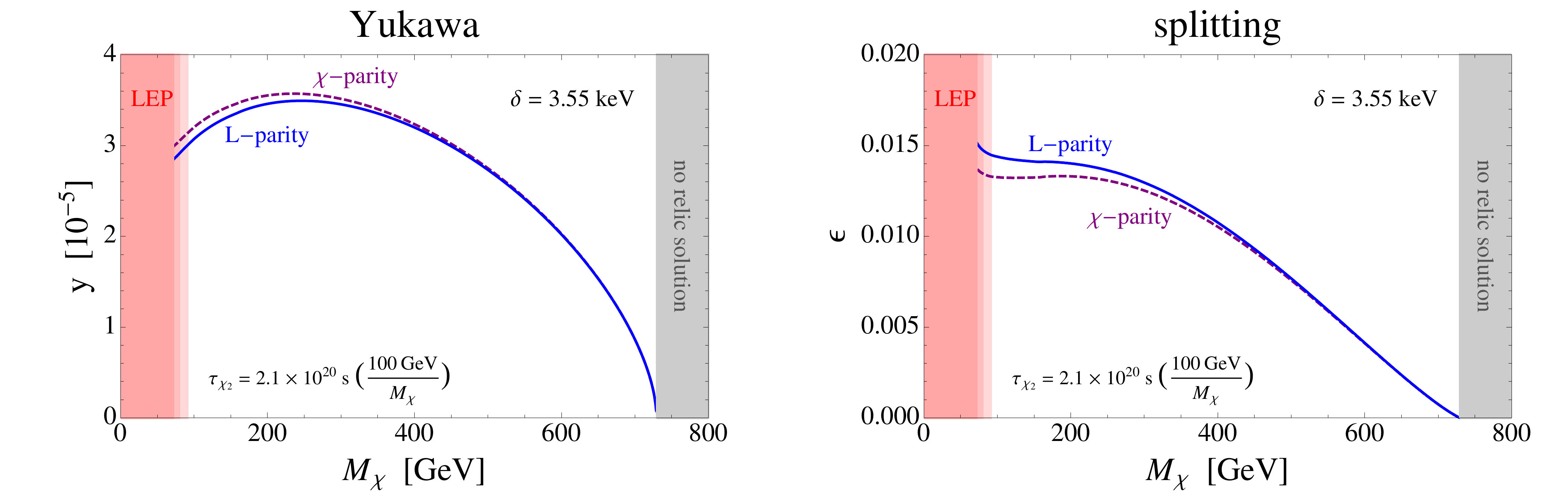}
\vspace*{-2mm}
          \caption{\footnotesize
{\bf Left:}  Magnitude of the Yukawa coupling $y$ as a function of $M_\chi$ required to fit the $3.55$~keV  X-ray line, with  $M_L$ fixed to reproduce the correct dark matter relic abundance. The solid line assumes $\tilde y = \tilde \lambda =0$ (approximate $L$-parity), while the dashed line assumes $\tilde y=\lambda=0$ (approximate $\chi$-parity). The shaded region at low $M_\chi$ is accessible to LEP-2, and the shaded region at high $M_\chi$ is where there is no dark matter relic solution for $\chi$.
{\bf Right:}  With the same assumptions,  the relative splitting between the two Yukawa couplings, $\epsilon=y/\lambda -1$ or $\epsilon=y/\tilde \lambda -1$, for approximate $L$-parity (solid line) or approximate $\chi$-parity (dashed line), respectively.
   }
\label{fig:couplings}
\end{center}
\vspace*{-3mm}
\end{figure}

%------------------------------------------------------------------------------------------
\subsection{LHC Phenomenology} \label{sec:lhc}

Apart for the X-ray line, the setup does not predict additional astrophysical signals.
The reason is that the mixing of dark matter with the active states needs to be small to accommodate observations. The resulting annihilation cross section in our galaxy is too small to observe. Likewise, this setup evades direct detection: the predicted spin-independent cross section, mediated by the inelastic $Z$-exchange between $\chi_1$ and $\chi_2$, is suppressed by two mixing angles and is thus at most of order $\sigma_{\rm DD}\sim 10^{-55}\, {\rm cm}^2$, well below the irreducible neutrino background.
However, our model does predict new phenomena that can be probed at the LHC\@.
The active vector-like pair $L$, $L^c$ couples to the electroweak gauge bosons of the SM, and therefore it can be produced in  proton-proton collisions.
For the range of masses consistent with  the dark matter abundance, $M_L \lesssim 700$~GeV, the production cross section is large enough to be accessible in the coming LHC run. In addition, for $M_L \lesssim 100$~GeV the active pair could have been produced at LEP-2.
Below we discuss the signatures that should be targeted in collider searches.

The charged and neutral components of the vector-like doublet can be produced in colliders via Drell-Yan processes.
Thanks to the mixing between $N$ and $\chi$, these particles can subsequently decay to dark matter.
The relevant  couplings are inherited from the gauge interactions of $L$:
\beq
s_{ij} {\sqrt{g_L^2 + g_Y^2} \over  2} Z^\mu \bar \chi_i \bar \sigma_\mu N_j
+{g_L \over 2 }  W_\mu^+ \left [ (s_{i1} + s_{i2}) \bar \chi_i  \bar \sigma_\mu E     +   (s_{i1} - s_{i2}) \chi_i \sigma_\mu  \bar E^c   \right ]
+ \hc,
\eeq
where $s_{ij}$ are the mixing angles defined in \eref{sij}.
Once produced,
$E$, $N_1$ and $N_2$ can thus  decay to dark matter $\chi_{1,2}$ and a $W$ or $Z$ boson.
Because the mass difference $\Delta = M_L - M_\chi$ is of order 10~GeV in the interesting parameter space, the vector  boson is off-shell and the decay is a 3-body one.
At the leading order in $\Delta$ the decay widths are given by:
\bea
\label{eq:Ndecay}
\sum_{i,f} \Gamma(E \to \chi_i f \bar f')  &\approx&
{ \Delta^5 \over 60 \pi^3 v^4} \left (s_{11}^2 + s_{22}^2  + 3 s_{12}^2 + 3 s_{21}^2 \right ) \sum_f  N_f ,
\nn
\sum_{i,f} \Gamma(N_1 \to \chi_i f \bar f)  &\approx&  {\Delta^5 \over 30 \pi^3 v^4 }
\left (s_{11}^2  + 3 s_{21}^2 \right)  \sum_f  N_f \left [ (T^3_f - Q_f \sin^2 \theta_W)^2 +Q_f^2 \sin^4 \theta_W \right ]\,,
\nn
\sum_{i,f} \Gamma(N_2 \to \chi_i f \bar f)  &\approx&  {\Delta^5 \over 30 \pi^3 v^4 }
\left (s_{22}^2  + 3 s_{12}^2 \right)  \sum_f  N_f \left [ (T^3_f - Q_f \sin^2 \theta_W)^2 +Q_f^2 \sin^4 \theta_W \right ]\,,
\eea
where $N_f$, $T^3_f$ and $Q_f$ are the number of colors, the isospin, and the electric charge of the SM fermion $f$ in the final state.
The range of the sum over  $f$ depends on $\Delta$: typically it runs over all  leptons, and the first two generations of quarks.
Recall that, with the simplifying assumption that only two of the four Yukawa couplings are non-zero, the only free parameter is the dark matter mass $M_\chi$.
It follows that the decay widths of $E$, $N_1$ and $N_2$ are then uniquely fixed for a given $M_\chi$.

 \begin{figure}[tb]
    \begin{center}
         \includegraphics[width=0.7\textwidth]{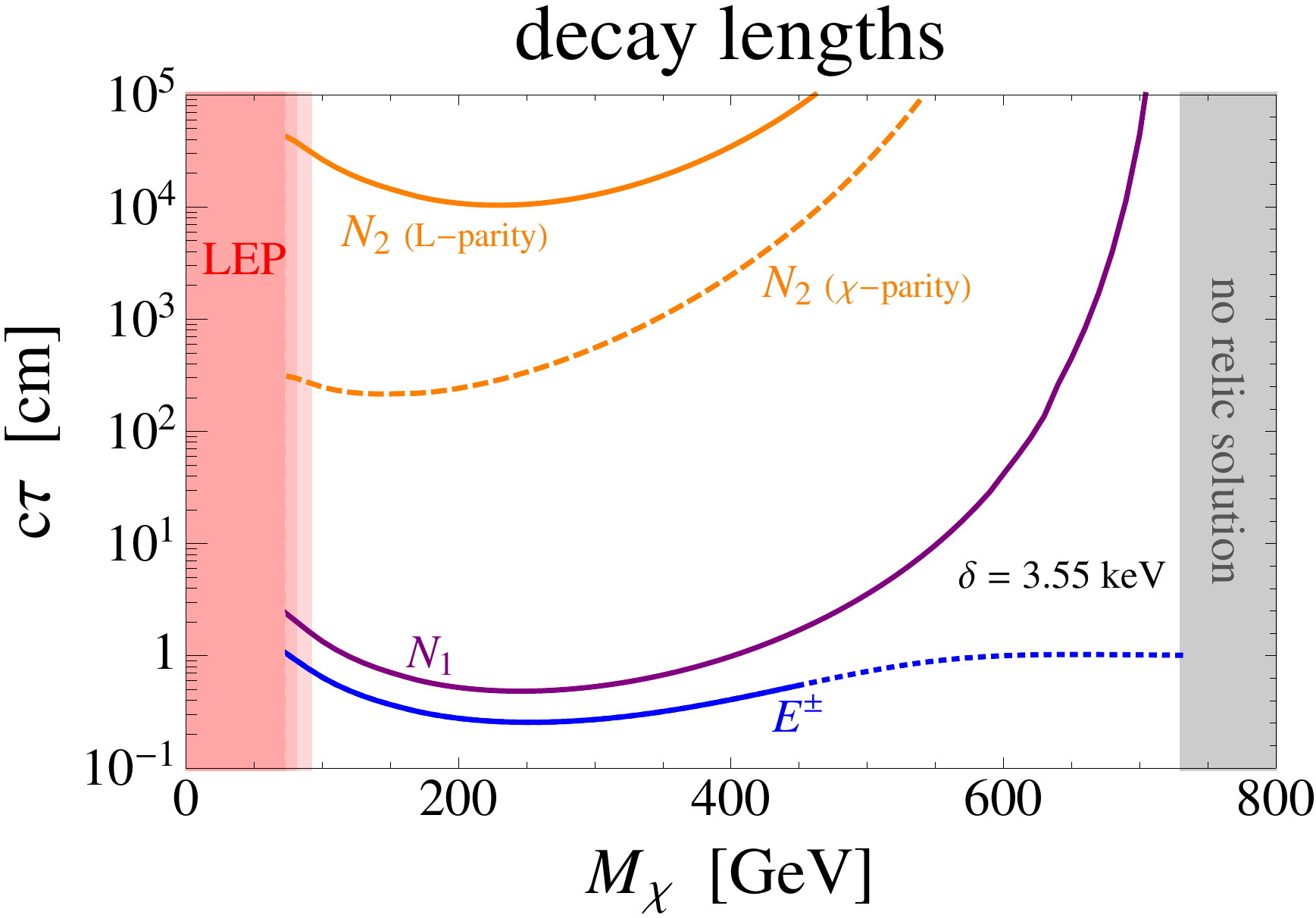}
\vspace*{-2mm}
          \caption{\footnotesize
Decay length of the active mass eigenstates $E$, $N_1$, $N_2$ (solid blue, solid purple, solid orange; bottom to top) for $\tilde y = \tilde \lambda = 0$ (approximate $L$-parity).
The other two Yukawa couplings are fixed to reproduce the observed 3.55~keV line and flux, and the masses of $E$, $N_1$ and $N_2$ are fixed to reproduce the correct dark matter abundance for $\chi$.  For $M_\chi \gtrsim 450$~GeV the $E \to N_i$ decays dominate over $E \to \chi_i$, which we indicate by the blue dotted line.
The dashed orange line is the $N_2$ decay length for  $\tilde y = \lambda = 0$ (approximate $\chi$-parity); in this case the $E$ and $N_1$ decay lengths are almost identical to the approximate $L$-parity case.
 The shaded region of $M_L \lesssim 100$~GeV is where the active states are accessible at LEP-2; the shaded region at high $M_L$ is where there is no dark matter relic solution for $\chi$.
   }
\label{fig:Ndecay}
\end{center}
\vspace*{-3mm}
\end{figure}

We plot the decay length in \fref{Ndecay}.
It is apparent that fitting the model to the observed 3.55~keV line leads to interesting LHC signatures.
For $\tilde y=\tilde \lambda=0$, where the keV line pushes towards the $L$-parity limit, we find that the decay length of $E$ and $N_1$ is
of ${\cal O}({\rm mm}-{\rm m})$ in our parameter space.
Thus, its production at the LHC may show up via displaced vertices in the detector, as long as $M_\chi \lesssim 700$~GeV\@.
For $M_\chi$ close to this upper limit, $N_1$ decays outside the tracker, or even outside the detector.
The coupling of $N_2$ to dark matter and $Z$ is suppressed by $\epsilon$
close to the $L$-parity limit, and thus its decay length is longer by a factor of $\sim 10^4$, which does not lead to displaced vertices.
The flattening in the decay length of $E$ is due to the weak decay $E \to N W^*$ winning over the decay to dark matter.
This decay is possible due to the splitting between $E$ and $N$ induced at one-loop by electromagnetic interactions, which  leads to
$m_E - m_N \sim 300$~MeV~\cite{Thomas:1998wy,Cirelli:2005uq}. Consequently, $E$ can decay to $N$ and (undetected) soft pions with the decay length of order one centimeter~\cite{Thomas:1998wy,Buckley:2009kv}.
For $\tilde y=\lambda=0$ where we are close to the $\chi$-parity limit, the situation is similar:
$N_1$ decays with displaced vertices for $M_\chi \lesssim 700$~GeV, and the lifetime of $N_2$ is larger by $\sim 10^4$.
The reason for the parametric difference of $N_2$ and $N_1$ decay lengths in this case is that
the coupling of $N_2$ to dark matter is suppressed by $\Delta/M_\chi \sim 10^{-1}-10^{-2}$ relative to that of $N_1$, leading to an enhanced lifetime. The decay length of $E$ and $N_1$ is similar near the $L$-parity and $\chi$-parity limits, since it is dominated by the decay into $\chi_1$, which exhibits a $\lambda \leftrightarrow \tilde \lambda$ symmetry between the two $\tilde y = \tilde \lambda=0$ and $\tilde y=\lambda=0$ cases.

The model thus predicts distinct collider signatures.
The Drell-Yan production of $E$ and $N_j$ and the decays $E \to \chi_i W^*$ and $N_j \to \chi_i Z^*$ lead to
events with missing energy and up to four soft non-collimated leptons or jets from leptonic $W$ or $Z$ decays, with vertices displaced from the interaction point.
The displacement depends on the parameter space and can vary between a millimeter and a meter.
This topology has not been explicitly searched for at LEP, the Tevatron and the LHC\@.
Therefore there is no strict lower limit on $M_L$, though we expect LEP to be sensitive to masses below $\sim 100$~GeV, indicated by the shaded region in Figs.~\ref{fig:coan},~\ref{fig:couplings} and~\ref{fig:Ndecay}.
The above signatures are challenging at the LHC, similar to probing nearly degenerate Higgsinos~\cite{Buckley:2009kv, Gori:2013ala, Baer:2014cua}, where search strategies include monojets and mono-photons.

We have surveyed the existing searches for displaced vertices at ATLAS~\cite{ATLAS:2012av, Aad:2012kw, TheATLAScollaboration:2013yia,Aad:2014yea} and CMS~\cite{CMS:2014mca,CMS:2013oea,CMS:2014bra}, and find that they have poor sensitivity for this topology.
In most analyses this is either due to a cut on a hard lepton or jet~\cite{TheATLAScollaboration:2013yia,CMS:2014mca,CMS:2013oea}, or the requirement of collimated leptons~\cite{Aad:2012kw,Aad:2014yea}.
In particular, in the recent CMS search for displaced supersymmetry in dilepton final states~\cite{CMS:2014bra}, the efficiency of the lepton momentum cuts for our signal is at the per mille level. As a result that search does not probe the parameter space of our model, and a more targeted analysis is in order. Likewise, the ATLAS analysis of~\cite{ATLAS:2012av} only excludes cross sections above $\sim1$~pb for lifetimes longer than $\sim0.5$~m. However, our model only has lifetimes longer than $\sim0.5$~m for masses above $\sim 500$~GeV where the cross section is smaller than $\sim5$~fb.
Detection prospects for nearly degenerate Higgsinos at future colliders such as the ILC and at a 100 TeV collider have also been studied~\cite{Berggren:2013vfa, Low:2014cba}, where the reach is improved. The decay process of our model has the additional handle of displaced decays through off-shell $Z$'s and $W$'s. Events with monojet/mono-photon + low-$p_T$ displaced leptons/jets thus represent a promising search strategy. We leave a detailed study of such a search for future work~\cite{future}.

%%%%%%%%%%%%%%%%%%%%%%%%%%%%%%%%%%%%%%%%%%%
\section{General X-ray lines}
\label{sec:anyother}

It is conceivable that new X-ray lines that cannot be matched to known atomic transitions will be discovered in current or future experiments.
In this section we consider the parameter space of our model that can accommodate general X-ray lines. This approach, akin to the simplified model approach to LHC phenomenology, offers a theoretical framework to describe such lines and predict possible correlations with other astrophysical and collider observables.
In particular, we are interested in identifying the range of line energies which can be accommodated within our model, the regions that are currently probed by X-ray experiments, as well as the resulting generic LHC features in the relevant parameter space.

For the following discussion we assume that only two Yukawa couplings, $y$ and either $\lambda$ or $\tilde \lambda$, are simultaneously non-zero and positive, and that $y$ is the larger of the two. We do not find qualitatively new effects if these assumptions are relaxed.
As before, we assume that co-annihilation between $\chi$ and $L$ is responsible for the observed dark matter abundance, which fixes
$M_L - M_\chi$  as a function of the dark matter mass.
In the general parameter space there are two additional constraints:
The first is that the lifetime of $\chi_2 \to \chi_1$, summing over the radiative and 3-body decays, should be longer than the age of the universe;
the second comes from existing experimental searches for new monochromatic lines in the X-ray spectrum.

\begin{figure}[tb]
    \begin{center}
         \includegraphics[width= \textwidth]{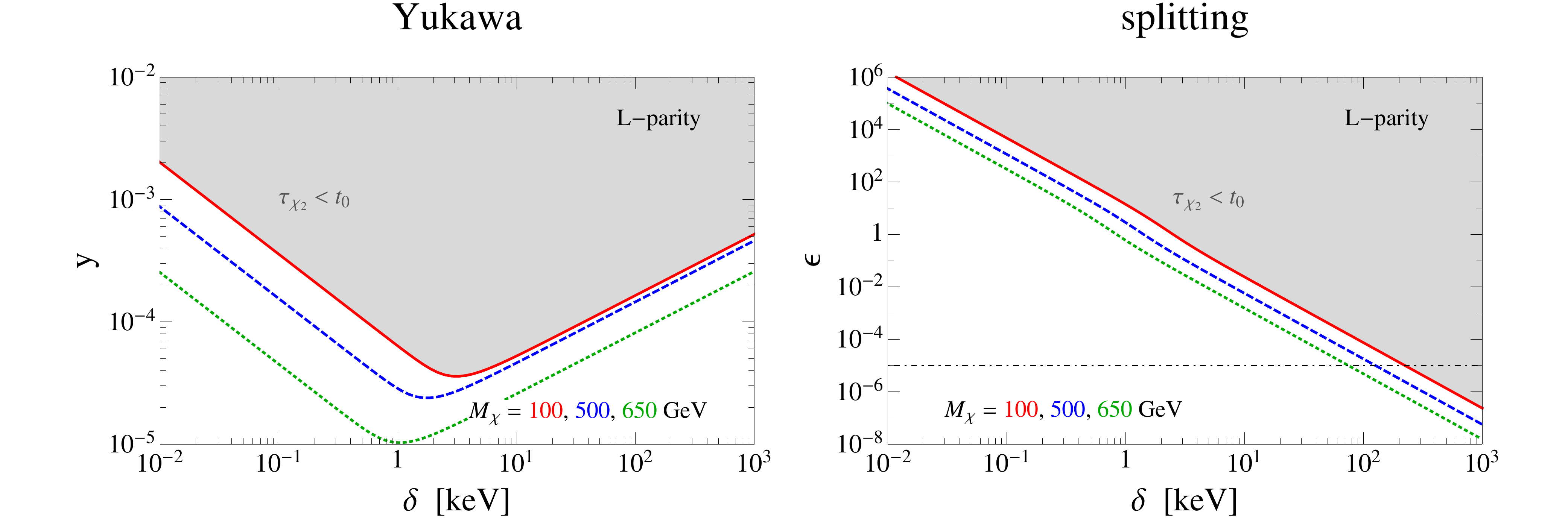}
         \vspace*{-2mm}
          \caption{\footnotesize
The Yukawa coupling $y$ (left panel) and splitting between the Yukawa couplings $\epsilon$ (right panel) that give  $m_{\chi_2} - m_{\chi_1} = \delta$, for dark matter mass $M_\chi=100,\ 500,\ 650$~GeV (solid red, dashed blue, dotted green) with the $\chi_2\to\chi_1$ decay time of $t_0=13.8\times 10^9$~years. Above the curves, the lifetime of $\chi_2$ is shorter than the age of the universe; the region below the curves is where galactic X-ray and gamma-ray lines can reside. In this plot we assume $\tilde y = \tilde \lambda = 0$ (approximate $L$-parity); the black dot-dashed curve indicates the naturalness lower bound on $\epsilon$. Similar results are obtained for $\tilde y=\lambda=0$ (approximate $\chi$-parity); see the text for details.
   }
\label{fig:yukawas_general}
\end{center}
\vspace*{-3mm}
\end{figure}

Concerning the dark matter lifetime, there are two relevant regimes, depending on the splitting $\delta$. Recall that for a given $\delta$, the product of the two Yukawa couplings is fixed, see~\eref{delta}.
For $\delta \gtrsim 1$~keV, to increase $\delta$ one needs to increase the magnitude of both Yukawa couplings.
At the same time, the difference between them must be made smaller in order to keep $\tau(\chi_2 \to \chi_1)$ longer than the age of the universe.
This means that either the $L$-parity or the $\chi$-parity limit must be approached.
In the case of $L$-parity, naturalness requires $\epsilon\gtrsim 10^{-5}$
due to the breaking of $L$-parity at 2-loop by the couplings of the hypercharge gauge boson to the SM fermions, setting
the naturalness limit $\delta \lesssim 100$~keV within our model.
In the case of $\chi$-parity, which is not broken by the SM sector, $\epsilon$ can be arbitrarily small in a technically natural way.
However, close to the $\chi$-parity limit, the mixing of $\chi_2$ and the active states is suppressed, and $\chi_2$ drops out of kinetic equilibrium with $L$ before the latter freezes out.
This violates the premises for co-annihilation, and as a consequence $\chi_2$ overcloses the universe.
As a result it impossible to accommodate $\delta$ larger than a few hundred keV in our model.
The setup therefore predicts an upper limit $\delta \lesssim 500$~keV on the possible energy of a monochromatic line;
this limit persists when  all four  Yukawa couplings are allowed to be non-zero, barring unnatural Yukawa couplings in the $L$-parity case.

For $\delta \lesssim 1$~keV, decreasing $\delta$ requires the product of the two Yukawa couplings to decrease, but they must be hierarchical rather than degenerate in order to keep $\tau(\chi_2 \to \chi_1)$  longer than the age of the universe.
In addition, ensuring that $\chi_2$ is kept in kinetic equilibrium with $L$ during co-annihilation requires one of the Yukawa couplings to be larger than $\sim10^{-5}$ in this regime of $\delta$.
Since the Yukawa couplings can be arbitrarily small without conflicting naturalness, we do not find a lower limit on the energy of a monochromatic line.
In principle, at very small $\delta$ when Yukawa couplings become sizable, limits from direct detection could become relevant. However, this is not the case in the parameter space relevant for X-ray satellites.
 For example, the current LUX limits~\cite{Akerib:2013tjd} can only exclude $\delta \lesssim 5$~eV for lifetimes equal to the age of the universe.
 For $\delta \gtrsim 100$~eV, the direct detection cross section is always below the level of the coherent neutrino background.

In \fref{yukawas_general} we show the size of the Yukawa couplings and their splitting, required to obtain the energy line $\delta$, for a variety of dark matter masses $M_\chi$. The depicted curves give $\tau(\chi_2\to \chi_1)$ of the age of the universe, and as such represent the upper limit on the Yukawa couplings that could give rise to observable galactic X-ray and gamma-ray lines. We find that the 3-body decay is negligible compared to the two-body one in the entire parameter space for X-rays and gamma-rays. The results are similar for the two cases $\tilde y=\tilde \lambda=0$ and $\tilde y=\lambda=0$ due to the near $\lambda \leftrightarrow \tilde \lambda$ symmetry between them (with respect to the decay rate and mass splitting $\delta$); thus only the former case is displayed. The kink in the left panel of \fref{yukawas_general} appears as one switches from the hierarchical to the degenerate coupling regime; see the discussion above.

\begin{figure}[tb]
    \begin{center}
    \includegraphics[width=\textwidth]{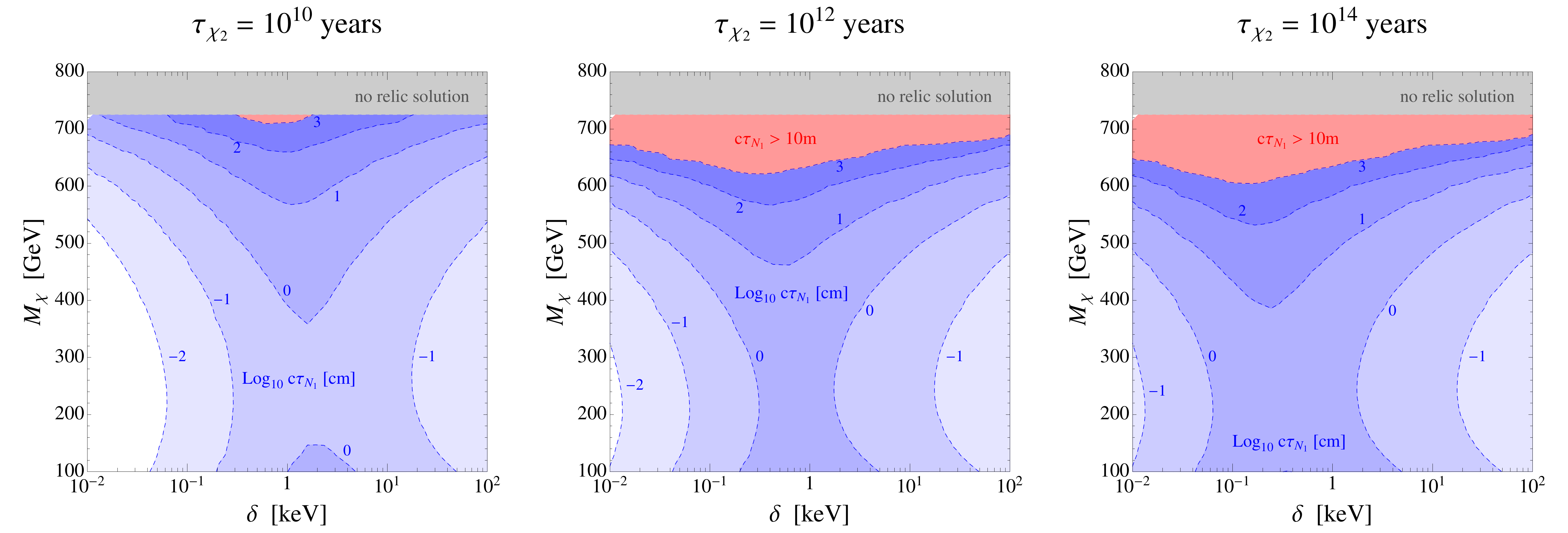}
         \vspace*{-2mm}
          \caption{\footnotesize
Regions in the $\delta$-$M_\chi$ parameter space where the decay width $\Gamma(N_1 \to \chi Z^*)$, summed over both dark matter states, corresponds to displaced vertices at colliders.
We assume $\tilde y  = \tilde \lambda = 0$ (approximate $L$-parity), and present results for three different dark matter lifetimes: $\tau(\chi_2 \to \chi_1) = 10^{10},10^{12},10^{14}$~years (left to right).
 For each $M_\chi$,  $M_L$ is fixed to reproduce the correct dark matter relic abundance.
     The various blue shaded regions correspond to $c\tau(N_1 \to \chi Z^*)$ between $0.1$~mm to 10~m. The white and pink regions correspond to $c\tau(N_1 \to \chi Z^*) < 0.1$~mm and $c\tau(N_1 \to \chi Z^*) > 10$~m, respectively. In the gray region there is no relic solution for $\chi$. Similar results are obtained for $\tilde y =\lambda=0$ (approximate $\chi$-parity).
   }
\label{fig:displacement_general}
\end{center}
\vspace*{-3mm}
\end{figure}

In \fref{displacement_general} we show the parameter space with $\tilde y  = \tilde \lambda = 0$, where decays $N_1 \to \chi Z^*$ (summed over both dark matter states) are displaced on the detector scale, for several dark matter lifetimes. Similar results are obtained for $\tilde y = \lambda =0$.
 The  decay length of $E  \to \chi W^*$ is always of the same order as that of $N_1 \to \chi Z^*$.
 We see that displaced vertices are associated not only with the 3.55~keV line, but are a typical characteristic of the model in the parameter space leading to X-ray lines.
Decays in the tracker, in the calorimeter, or in the muon chambers can be realized, leading to a wide spectrum of possible displaced signals at the LHC\@.
For $\delta \lesssim 1$~keV, when the Yukawa couplings $y$ and $\lambda$ are hierarchical, the displacement strongly depends on the dark matter lifetime.
In this region both $N_1$ and $N_2$ have similar decay length, as both are set by the magnitude of the Yukawa coupling $y$.
For $\delta \gtrsim 1$~keV, when the Yukawa couplings are close to the $L$-parity limit $y \approx \lambda$, the decay length of $N_1$ is weakly dependent on the dark matter lifetime. In this region the decay width of $N_2$ is suppressed by the small splitting between the Yukawa couplings, leading to decays outside the detector. For large masses $M_\chi$, corresponding to small mass splitting $\Delta$, $N_1$ can decay outside the detector as well.

In \fref{spaceL} we show the range of $\chi_2$ lifetime available in our model close to the $L$-parity (upper plots) and $\chi$-parity (lower plots) limits, along with the exclusion limits from X-ray experiments. We plot the results for $M_\chi=100$~GeV (left plots) and $650$~GeV (right plots); for masses in between these values the results fall between the depicted ones. The blue solid [dashed] curve in the upper panel of~\fref{spaceL} indicates the naturalness limit for the Yukawa couplings, $\epsilon\gtrsim 10^{-4}\ [10^{-5}]$
-- for lifetimes exceeding this curve, a tuning is needed due to the radiative breaking of $L$-parity. The shaded blue region in the lower panel of~\fref{spaceL} indicates
where $\chi_2$ falls out of equilibrium and co-annihilation cannot persist. The bottom shaded regions in all panels indicate the exclusion limits from current X-ray experiments: Chandra~\cite{Chandra} observations of the Andromeda galaxy~\cite{Watson:2011dw} (shaded green); HEAO-1 and XMM-Newton observations of the cosmic X-ray background (CXB)~\cite{Boyarsky:2005us}  (shaded red); the most stringent constraints~\cite{Watson:2006qb, Watson:2011dw} from nearby galaxies and clusters~\cite{Abazajian:2006yn, Abazajian:2005xn, Boyarsky:2006zi, Boyarsky:2006fg, Boyarsky:2006ag, Abazajian:2006jc} (shaded purple); and Integral measurements of the unresolved X-ray emission from the Milky Way halo~\cite{Bouchet:2008rp, Essig:2013goa} (shaded orange). Thus, the parameter space above the bottom filled regions and below the blue shaded region remains to be explored in future X-ray experiments.

\begin{figure}[tb]
    \begin{center}
         \includegraphics[width=0.99\textwidth]{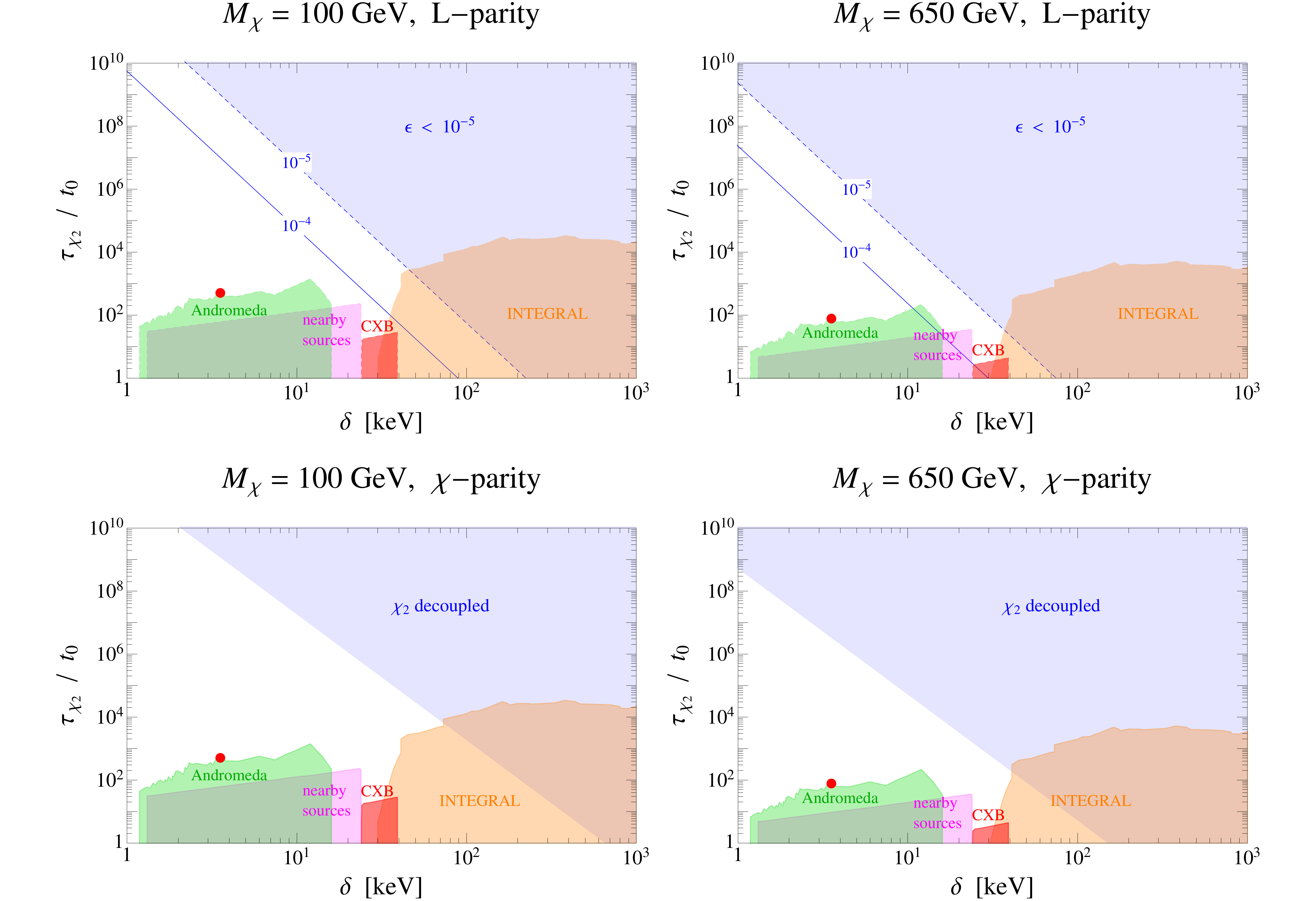}
\vspace*{-2mm}
          \caption{\footnotesize
The available lifetime in our model, in units of the present age of the universe $t_0$, as a function of the line energy, for $M_\chi = 100$~GeV (left), and for $M_\chi = 650$~GeV (right).
The top plots assume $\tilde y = \tilde \lambda =0$ (approximate $L$-parity), and the solid [dashed] blue curve indicates the naturalness bound $\epsilon=y/\lambda-1\gtrsim 10^{-4}$ [$10^{-5}$] due to radiative breaking of $L$-parity.
The bottom plots assume $\tilde y =  \lambda =0$ (approximate $\chi$-parity), and the shaded blue region indicates where $\chi_2$ falls out of equilibrium and co-annihilations cannot proceed.
The bottom shaded regions in all plots indicate the exclusion limits from current X-ray experiments: Chandra~\cite{Chandra} observations of the Andromeda galaxy~\cite{Watson:2011dw} (shaded green); HEAO-1 and XMM-Newton observations of the cosmic X-ray background (CXB)~\cite{Boyarsky:2005us}  (shaded red); the most stringent constraints~\cite{Watson:2006qb, Watson:2011dw} from nearby galaxies and clusters~\cite{Abazajian:2006yn, Abazajian:2005xn, Boyarsky:2006zi, Boyarsky:2006fg, Boyarsky:2006ag, Abazajian:2006jc} (shaded purple); and Integral measurements of the unresolved X-ray emission from the Milky Way halo~\cite{Bouchet:2008rp, Essig:2013goa} (shaded orange).
The range below the blue shaded region and above the bottom shaded regions is the natural and allowed parameter space for future X-ray experiments.
   }
\label{fig:spaceL}
\end{center}
\vspace*{-3mm}
\end{figure}

%%%%%%%%%%%%%%%%%%%%%%%%%%%%%%%%%%%%%%%%%
\section{Summary} \label{sec:conclusions}

In this paper we discussed the possibility of an X-ray line arising from thermal weak-scale dark matter.
In the spirit of simplified models, we presented a minimal module where such a scenario can be realized.
Dark matter consists of two nearly degenerate pseudo-Dirac states with a mass splitting due to small Majorana masses.
Moreover, we introduced a doublet of active states charged under the SM which provided the portal between dark matter and the visible sector.
The dark matter abundance is populated thermally in the early universe via co-annihilation with the active states.
The X-ray line arises from the decay of the heavier dark matter component into the lighter one via a radiative dipole transition, at a rate that is slow compared to the age of the universe.
This model can explain the recently observed $3.55$~keV X-ray line, but more generally it serves as a benchmark framework for interpretations of results from X-ray satellites.
At the LHC, the model predicts exotic events with up to four soft non-collimated displaced leptons or jets and missing energy.
This motivates new signatures at the LHC that can be searched for.

%%%%%%%%%%%%%%%%%%%%%%%%%%%%%%%%%%%%%%
\section*{Acknowledgements}

We thank Eric Kuflik and Neal Weiner for helpful conversations, and Howie Haber for useful discussions and for a secret version of the spinor bible.
AF is supported by the ERC advanced grant Higgs@LHC\@. The work of YH is supported by the U.S. National Science Foundation under Grant No. PHY-1002399. YH is an Awardee of the Weizmann Institute of Science - National Postdoctoral Award Program for Advancing Women in Science.  J.T.R. was supported in part by the Miller Institute for Basic Research in Science.

\bibliographystyle{JHEP}
\bibliography{kevlinepaper}

\end{document}